\documentclass[
superscriptaddress,
longbibliography,
twocolumn,
aps,prl]{revtex4-2}

\usepackage{graphicx}
\usepackage{dcolumn}
\usepackage{bm}
\usepackage{xcolor}
\usepackage[mathlines]{lineno}
\usepackage[normalem]{ulem}
\usepackage{epstopdf}
\usepackage[
			colorlinks = true,
            linkcolor = blue,
            urlcolor  = blue,
            citecolor = blue,
            anchorcolor = blue
            ]{hyperref}         
\newcommand\rxout{\bgroup\markoverwith{\textcolor{red}{\rule[.5ex]{2pt}{.6pt}}}\ULon}

\begin{document}

\title{Spatiotemporal Vortex Rings Induced by Spatiotemporal Coupling}

\author{Zhiling Zhou}
\thanks{Z. Z. and W. Z. contributed equally to this work.}
\affiliation{Department of Physics, City University of Hong Kong, Kowloon, Hong Kong, China}

\author{Wei Zhong}
\thanks{Z. Z. and W. Z. contributed equally to this work.}
\affiliation{Institute of Acoustics, Tongji University, Shanghai 200092, China}

\author{Tong Fu}
\affiliation{Department of Physics, City University of Hong Kong, Kowloon, Hong Kong, China}

\author{Wanyue Xiao}
\affiliation{Department of Physics, City University of Hong Kong, Kowloon, Hong Kong, China}

\author{Zhongming Gu}
\affiliation{Institute of Acoustics, Tongji University, Shanghai 200092, China}

\author{Jie Zhu}
\email{jiezhu@tongji.edu.cn}
\affiliation{Institute of Acoustics, Tongji University, Shanghai 200092, China}

\author{Shubo Wang}
\email{shubwang@cityu.edu.hk}
\affiliation{Department of Physics, City University of Hong Kong, Kowloon, Hong Kong, China}

\date{\today}

\begin{abstract}
Vortices and vortex rings are topological structures that arise in various physical systems. However, the generation of spatiotemporal vortices (STVs) and vortex rings (STVRs) has so far relied on complex, often active wavefront modulation. We theoretically and experimentally demonstrate that spatiotemporal coupling can drive unstructured wave packets to form vortices upon scattering from simple obstacles. The resulting STVs and STVRs possess controllable topological charges and excellent propagation stability. These findings reveal a fundamental mechanism for spatiotemporal singularity formation and provide a universal route to structured-wave generation.
\end{abstract}

\maketitle
\textit{Introduction.\textemdash}Vortices and vortex rings are intriguing topological structures that arise in a wide range of physical systems \cite{NakayamaN_2018Nature,YacoubMH_2000Nature}, such as fluid vortices induced by vortex-sheet roll-up \cite{akhmetov2009vortex}, quantum vortices in Bose–Einstein condensates \cite{Brian_2008Nature}, and magnetic vortices in ferromagnets \cite{IvanK_2010PRB}. In classical wave systems, beyond monochromatic spatial vortices \cite{WoerdmanJP_1992PRA,RicardoR_2008PRL,ChanCT_2018SA,WangYS_2020APL,MaG_2021NC,StevenC_2021CP,xiao2026acoustic}, polychromatic variants can also emerge, known as spatiotemporal vortices (STVs) \cite{BerryM_1997PRSL,MilchbergHM_2016PRX,RuanZ_2022LPR,CTChan_2024NL,DaiYN_2026PRL,XuT_2024NC,ChenYF_2025NC,ChenYF_2023PRL,MilchbergHM_2019optica} and spatiotemporal vortex rings (STVRs) \cite{ZhanQ_2022NPho,CaiY_2025SA,ShenY_2025SA}. Unlike spatial vortices carrying longitudinal orbital angular momentum (OAM), STVs and STVRs possess transverse OAM and unique topological features \cite{BliokhKY_2012PRA,BliokhK_2023PRA}, which give rise to rich phenomena and profound applications including spin-orbit interactions \cite{Bliokh_2021PRL,ZhanQ_2022ACS}, OAM-based information encoding \cite{WanC_2025ACSPhotonics,ZhanQW_2024NC,YaoJP_2024SA}, and structured-wave manipulation \cite{MarkR_2021NPho,ShenY_2024APR,MK_2025NC,NI_2022NPho}. 

Due to their polychromatic nature, STVs and STVRs are inherently inseparable in space and time. Spatiotemporal coupling is ubiquitous throughout their formation and evolution. Such coupling, however, has traditionally been regarded as a source of distortion and other detrimental effects  \cite{akturk2010spatio,Fabien_2016NPh,NiekF_2011_OE}. Consequently, conventional generation schemes rarely exploit this coupling, relying instead on complex phase engineering using sophisticated devices like spatial light modulators, resonant gratings, or active phased arrays \cite{RuanZ_2023NC,LiaoCT_2021NPho,ZiJ_2024PRL,ZhanQ_2022NPho,WuDJ_2026PRapplied}. Furthermore, the resulting STVs and STVRs generally suffer from the diffraction-dispersion imbalance that rapidly destroys the vortex core upon propagation \cite{MilchbergHM_2021PRL,ZiJ_2025SB,ZhanQ_2022SB}, presenting a fundamental challenge to preserving these topological wave fields.

In this Letter, we theoretically and experimentally demonstrate that spatiotemporal coupling can actually serve as a constructive mechanism for vortex formation. This coupling can reshape topologically trivial wave packets into stable STVs and STVRs. Remarkably, their topological charges remain well preserved over long-distance propagation, despite the inherent diffraction-dispersion imbalance. Unlike conventional highly engineered methods, the proposed mechanism arises naturally when wave packets scatter off simple obstacles, offering an elegant yet robust route to generating stable STVs and STVRs.

\begin{figure}[b]
\centering
\includegraphics[width=8.8cm]{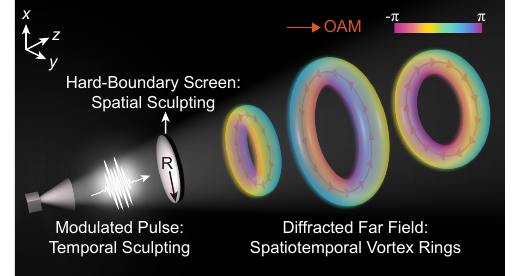}
\caption{Schematic of STVRs induced by spatiotemporal coupling. A Gaussian wave packet is spatially (by the screen) and temporally (by the time modulation) sculpted, giving rise to STVRs in the far field.}
\label{fig:1}
\end{figure}

\begin{figure}[t]
\centering
\includegraphics[width=8.8cm]{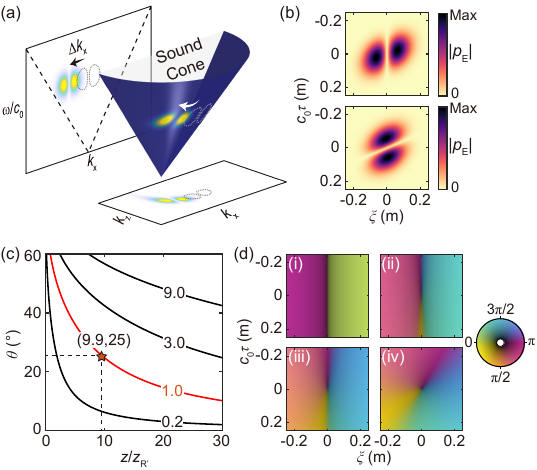}
\caption{Spatiotemporal coupling and vortex formation. (a) Schematic spectra of the wave packet on the sound cone, ($\omega$,${k_x}$) and (${k_x}$,${k_z}$) planes. The cone with ${k_z} < 0$ is omitted. The arrows indicate that the wave packet obtains a transverse momentum $\Delta k_x$ due to scattering. (b) Envelopes of the spatial dipole (upper panel) and the temporal dipole (lower panel) at $z=2{z_{{\rm{R'}}}}$ with $\theta  = {-25^\circ }$. (c) Contours of the ellipticity $\eta$ with respect to $\theta$ and $z$. (d) Normalized amplitudes and phases of the vortex core at $z=0$ (i), $0.8{z_{{\rm{R'}}}}$ (ii), $2{z_{{\rm{R'}}}}$ (iii), $9.9{z_{{\rm{R'}}}}$ (iv), with $\theta  = {-25^\circ }$. The panel (iv) corresponds to the star noted in (c). The results in (b-d) are calculated in the co-moving frame. The central frequency, central wavelength, spatial width, and temporal width of the wave packet are $f_{\rm{c}}=6000$ Hz, $\lambda_{\rm{c}}\approx 0.057$ m, ${\sigma _{\rm{\xi}}} = \lambda_{\rm{c}}/2$ and ${\sigma _\tau } = 1/{f_{\rm{c}}}$, respectively.  }
\label{fig:2}
\end{figure}

\begin{figure*}[t]
\centering
\includegraphics[width=17.6cm]{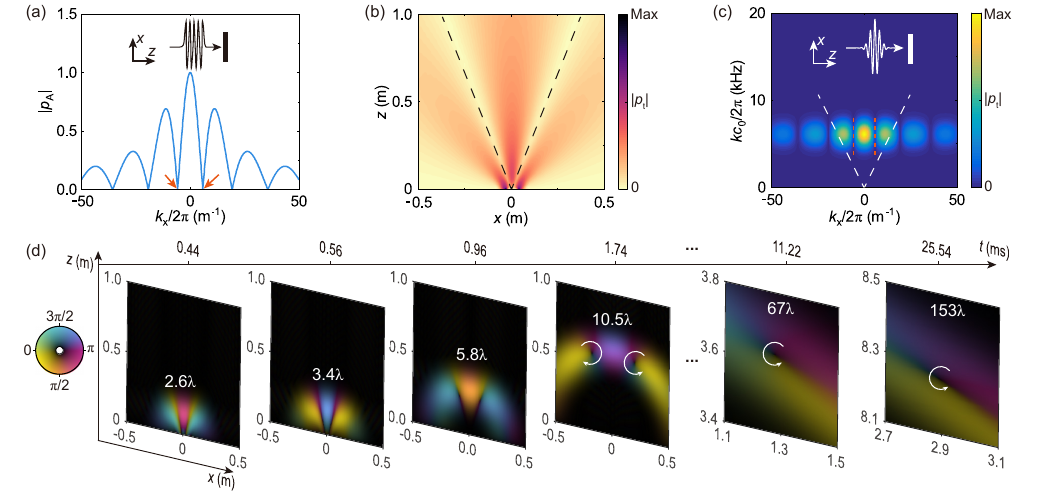}
\caption{Realization of 2D STVs. (a) Normalized angular spectrum for a Gaussian beam scattered by a screen of length $\lambda_c$. (b) Scattered field corresponding to the spectrum in (a). (c) Normalized polychromatic angular spectrum for an incident Gaussian wave packet with spatial width ${\sigma _x} = \lambda_{\rm{c}}/2$ and temporal width ${\sigma _t } = 1/{f_{\rm{c}}}$. The white and red dashed lines depict the sound cone and the nodal lines, respectively. (d) Evolution of the transient scattered field. The field amplitude and phase are denoted by the brightness and color, respectively. The rapidly oscillating phase of the carrier is removed. The central frequency $f_{\rm{c}}$ and wavelength $\lambda_c$ are the same as those in Fig.~\ref{fig:2}. The whole evolution process in (c) is provided in Supplementary Video 1.} 
\label{fig:3}
\end{figure*}

\begin{figure}[t]
\centering
\includegraphics[width=8.8cm]{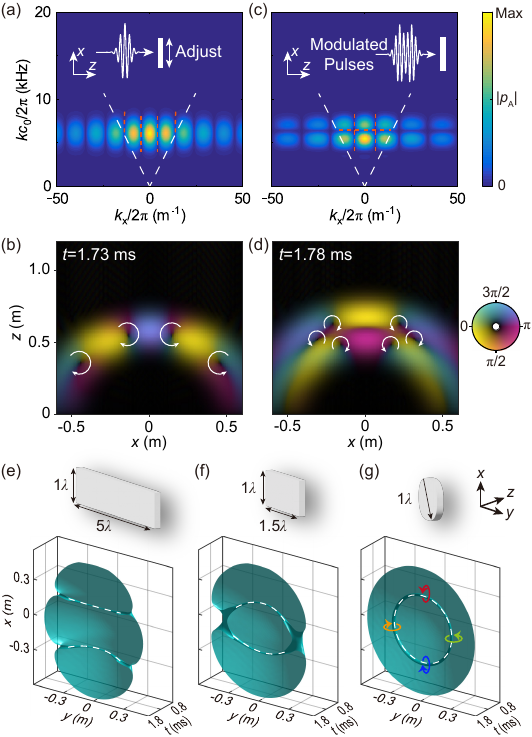}
\caption{Manipulation of the topological charges and STVR formation. (a) Normalized polychromatic angular spectrum for a Gaussian wave packet (${\sigma _t} = 1/f_{\rm{c}}$, ${\sigma _x} = \lambda_{\rm{c}}/2$) scattered by the plate of length $1.6\lambda_{\rm{c}}$. (b) Scattered field corresponding to the spectrum in (a) at $t=1.73$ ms. The arrows depict the winding direction of the vortex. (c) Normalized polychromatic angular spectrum for a modulated wave packet scattered by the plate of length $\lambda_{\rm{c}}$. The modulated wave packet is the superposition of two Gaussian pulses (${\sigma _t} = 1/f_{\rm{c}}$, ${\sigma _x} = \lambda_{\rm{c}}/2$) with a time interval $\Delta t = 2.4/f_{\rm{c}}$. (d) Scattered field corresponding to the spectrum in (c) at $t=1.78$ ms. (e-f) Evolution of singular lines in the spatiotemporal domain as the plate varies from a rectangular slab to a disc. The surfaces correspond to the iso-amplitude surfaces ($\left| p_{\rm{t}} \right| = 0.02$) of the scattered fields at $z=0.5$ m. The spatial width and temporal width of the wave packet in (e-g) are ${\sigma _r} = \lambda_{\rm{c}}/2$ and ${\sigma _t} = 1/f_{\rm{c}}$. The central frequency $f_{\rm{c}}$ and wavelength $\lambda_c$ are the same as those in Fig.~\ref{fig:2}. The whole evolution processes in (b,d) are provided in Supplementary Video 2 and 3. } 
\label{fig:4}
\end{figure}

\begin{figure*}[t]
\centering
\includegraphics[width=17.6cm]{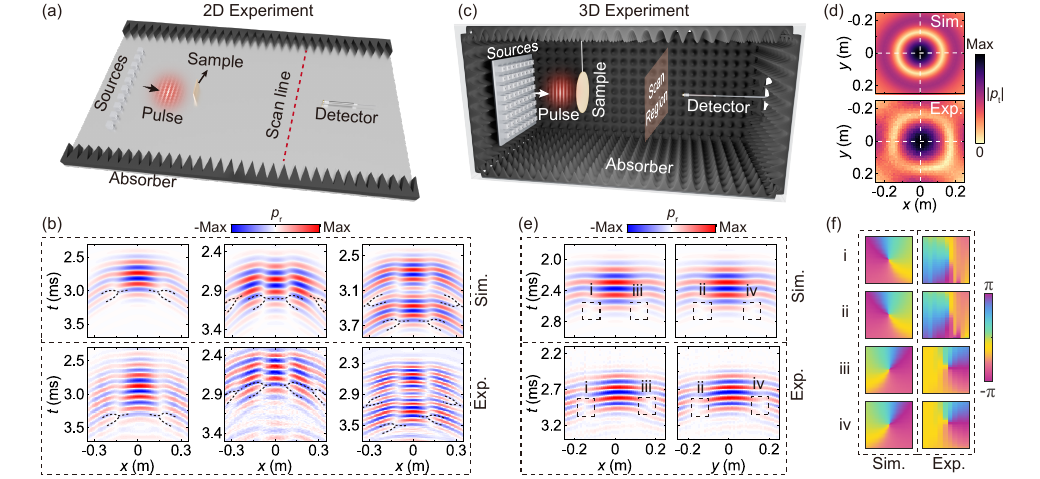}
\caption{Experimental verification of the STV and STVR generation. (a) Schematic of the 2D experiment platform. (b) Scattered fields in the spatiotemporal domain along the scan line marked in (a). The dashed lines depict the shapes of the dislocations. The first and second columns correspond to samples with lengths of $25$ mm and $125$ mm, respectively, where the pulse parameters are $f_{\rm{c}}=6000$ Hz and ${\sigma _t} = 1/f_{\rm{c}}$. The third column corresponds to the sample with length of $75$ mm, where the incident pulse is the superposition of two pulses ($f_{\rm{c}}=6000$ Hz, ${\sigma _t} = 1/f_{\rm{c}}$) separated by a time interval $\Delta t = 3/f_{\rm{c}}$. (c) Schematic of the 3D experiment platform. (d) Amplitude of the scattered fields in space domain within the scan region in (c). The simulated and experimental results correspond to $t=2.6$ ms and $2.9$ ms, respectively. The pulse parameters are $f_{\rm{c}}=6000$ Hz, ${\sigma _t} = 1/f_{\rm{c}}$. The diameter of the sample is $100$ mm. (e) Scattered fields along the dashed lines in (d). (f) Phase profiles within the square regions in (e). ``Sim." denotes ``Simulation". ``Exp."  denotes ``Experiment".}
\label{fig:5}
\end{figure*}

\textit{Vortex formation mechanism.\textemdash}
We consider an incident Gaussian wave packet scatters off a hard-boundary disc, as shown in Fig.~\ref{fig:1}. We formulate the problem in $xoz$-plane due to the cylindrical symmetry of the system. The evolution of the wave packet after scattering is governed by the following wave equation in the co-moving frame (i.e., the frame of the wave packet) \cite{SI}
\begin{equation}
\label{eq:1}
i\frac{{\partial {{\tilde p}_{\rm{E}}}}}{{\partial z}} = \hat H{{\tilde p}_{\rm{E}}},
\end{equation}
where
\begin{equation}
\label{eq:2}
\hat H = \frac{1}{{2{k_{{\rm{eff}}}}}}\frac{{{\partial ^2}{}}}{{\partial {\xi ^2}}} + \gamma \frac{{{\partial ^2}{}}}{{\partial \xi \partial \tau }} + \frac{{\alpha \gamma }}{2}\frac{{{\partial ^2}{}}}{{\partial {\tau ^2}}}
\end{equation}
is the spatiotemporally-coupled propagator. Here, $\xi  = x - z\tan \theta $ and $\tau  = t - z/({c_0}\cos \theta )$ denote the co-moving space and time coordinates, respectively, ${k_{{\rm{eff}}}} = {k_0}{\cos ^3}\theta$ characterizes the diffraction strength, $\gamma  = \tan \theta /({c_0}{k_0}{\cos ^2}\theta )$ quantifies the spatiotemporal coupling strength, $\alpha  = \gamma {k_{{\rm{eff}}}}$ characterizes the angular dispersion, ${\tilde p}_{\rm{E}}$ is the wave packet envelope, $\theta$ is the angle between the wave propagation direction and $z$-axis.

The dispersion relation of the acoustic waves in air is ${k_z} = \sqrt {{k^2} - k_{x}^2}$, where $k$, $k_x$, and $k_z$ are the total, transverse and axial wavenumber, respectively, and $c_0=343$ m/s is the sound speed. This relation can be visualized as a conical surface in the momentum space, i.e., the sound cone [Fig.~\ref{fig:2}(a)]. The incident wave packet corresponds to a finite region on the sound cone with specific spatial and temporal frequency components \cite{Abouraddy_2022AOP}. In the near field, the screen sculpts the wave packet into multiple lobes separated by nodal lines via scattering. The lowest order component of the scattered wave packet has a dipole envelope with two lobes, as shown by the colormap on the sound cone and in the $\omega$-$k_x$ and $k_x$-$k_z$ spaces [Fig.~\ref{fig:2}(a)]. This dipole envelope can be expressed as ${\tilde p_{\rm{E}}}({\xi _0},0,\tau ) = {\xi _0}\exp ( { - {\xi _0}^2/\sigma _\xi ^2 - {\tau ^2}/\sigma _\tau ^2} )$, where ${\sigma _\xi }$ and ${\sigma _\tau }$ denote the spatial and temporal widths, respectively. The evolution of the dipole envelope can be determined using the propagator $\hat H$ in Eq.~(\ref{eq:2}), where space and time become inseparable due to spatiotemporal coupling. The obtained envelope in the sheared frame ($\xi ' = \xi ,{\rm{ }}\tau ' = \tau  - \alpha \xi $) is \cite{SI}
\begin{equation}
\label{eq:3}
{\tilde p'_{\rm{E}}}(\xi ',z,\tau ') = \frac{{\xi 'E(\xi ',z,\tau ')}}{{{D^{3/2}}(z)}} + i\frac{{\zeta (z)}}{{{c_0}\tau '}}\frac{{{c_0}\tau 'E(\xi ',z,\tau ')}}{{{D^{3/2}}(z)}},
\end{equation}
where $\zeta (z) = B(\tau ')z/{k_{{\rm{eff}}}}$, $B\left( {\tau '} \right) = \alpha \tau '/\sigma _\tau ^2$, $D(z) = 1 - iz/{z_{\rm{R}'}}$ includes the Gouy phase, with ${z_{{\rm{R'}}}} = {k_{{\rm{eff}}}}/[2(1/2\sigma _\xi ^2 + {\alpha ^2}/2\sigma _\tau ^2)]$, and $E \left( {\xi ',z,\tau '} \right)$ is an exponential function that describes the broadening of the envelope \cite{SI}. 

Equation~(\ref{eq:3}) comprises two terms with a phase difference of $\pi /2$. The first term corresponds to a spatial dipole $\xi 'E(\xi ',z,\tau ')/{D^{3/2}}(z)$ with an amplitude of $1$, while the second term corresponds to a temporal dipole ${c_0}\tau 'E(\xi ',z,\tau ')/{D^{3/2}}(z)$ with an amplitude of $\zeta (z)/{c_0}\tau '$. Figure~\ref{fig:2}(b) shows the envelopes of the spatial dipole (upper panel) and the temporal dipole (lower panel). Their superposition gives rise to a STV, similar to the vortex formed by the superposition of two orthogonal Hermite-Gaussian modes. Notably, the two dipoles in Eq.~(\ref{eq:3}) are nonorthogonal due to the spatiotemporal coupling, which is distinct to the orthogonal Hermite-Gaussian modes (see Ref. \cite{SI} for the details). The purity of this STV can be quantified by an ellipticity defined as the amplitude ratio of the two dipoles $\eta (z) = \zeta (z)/({c_0}\tau ')$. Figure~\ref{fig:2}(c) shows the contours of $\eta = 0.2, 1.0, 3.0$ and $9.0$ as a function of the propagation angle $\theta$ and propagation distance $z$. We note that the spatiotemporal coupling vanishes for $\theta  = {0^ \circ }$, in which case $\eta =0$, and the wave packet remains topologically trivial during propagation. The ellipticity gradually increases with $z$ for a fixed $\theta$. At a larger $\theta$, the spatiotemporal coupling strength $\gamma$ becomes more pronounced, thereby leading to a more dramatic variation in the ellipticity. When $\theta  = {-25^\circ }$, $\eta$ is close to unity at $z \approx 10{z_{{\rm{R'}}}}$ (marked by the star), indicating the emergence of a pure vortex. Figure 2(d) shows the normalized amplitude and phase profiles of the vortex core ${{[\xi ' + i\zeta (z)]} \mathord{\left/ {\vphantom {{[\xi ' + i\zeta (z)]} {{D^{3/2}}(z)}}} \right. \kern-\nulldelimiterspace} {{D^{3/2}}(z)}}$ for $z=0$, $0.8{z_{{\rm{R'}}}}$, $2{z_{{\rm{R'}}}}$, $9.9{z_{{\rm{R'}}}}$, which evolves from a dipole-like distribution to a vortex. These results clearly demonstrate how spatiotemporal coupling drives the initial state ${\tilde p_{\rm{E}}}({\xi _0},0,\tau )$ in Fig.~\ref{fig:2}(a) to evolve into a vortex through the propagator $\hat H$. Crucially, the spatial and temporal dipoles in Eq.~(\ref{eq:3}) share the same Gouy phase \cite{WinfulHG_2001OL} (included in $D\left( z \right)$), which protects the STV during propagation. In contrast, the conventional STVs generally suffer from an additional Gouy phase difference due to the unbalanced diffraction and dispersion \cite{ChenYF_2023PRL,MilchbergHM_2021PRL}, which destroys the vortex topological charge within a few Rayleigh lengths.

\textit{Propagation-stable STVs and STVRs.\textemdash}
To understand the spatial sculpting enabled by the screen, we first consider a monochromatic Gaussian beam incident on a thin plate with hard boundaries [inset in Fig.~\ref{fig:3}(a)], which is a two-dimensional (2D) system infinitely extended in $y$-direction (out of the page). The scattered field can be determined by using polychromatic angular spectrum theory (see Ref. \cite{SI} for the details). The angular spectrum of the scattered field, shown in Fig.~\ref{fig:3}(a), exhibits a series of zeros (i.e., nodes). The first-order zeros (marked by the red arrows) give rise to two nodal lines in the scattered field [dashed lines in Fig.~\ref{fig:3}(b)], corresponding to the spatial sculpting by the thin plate. Next, we consider a polychromatic Gaussian wave packet (the spectrum is Gaussian in the $\omega$-$k_x$ domain) scattered by the same plate [inset of Fig.~\ref{fig:3}(c)]. In this case, two nodal lines (denoted by the dashed red lines) emerge inside the sound cone in $\omega$-$k_x$ space [Fig.~\ref{fig:3}(c)], sculpting the wave packet into separated lobes. The spectral components outside the sound cone correspond to the evanescent fields that do not contribute to the vortex formation in the far field. The evolution of the scattered field is shown in Fig.~\ref{fig:3}(d). We observe that the wave packet exhibits a spherical wavefront at $L = 2.6 \lambda$. At $L = 3.4 \lambda$, the wave packet splits into three lobes separated by two nodal lines, consistent with the results in Fig.~\ref{fig:3}(c). As predicted by Eq.~(\ref{eq:3}), the spatiotemporally-coupled propagator $\hat H$ deforms the nodal lines and gradually builds up the vorticity during propagation. Eventually, two vortices carrying opposite topological charges emerge in the far field at $L = 10.5 \lambda$. These charges remain well preserved even after a propagation distance of over $L = 67\lambda $, demonstrating the excellent stability of the STV.

The number of vortices in the scattered wave packet can be easily controlled by adjusting the width of the plate. A larger width of the plate leads to more nodal lines inside the sound cone [Fig.~\ref{fig:4}(a)], and thereby more vortices are formed in the far field [Fig.~\ref{fig:4}(b)]. Besides, the sign of the topological charges can be reversed via temporal sculpting, i.e., modulating the time interval between two incident pulses. This gives rise to a new nodal line parallel to the $k_x$ axis in the spectrum, dividing the wave packet along the frequency dimension [Fig.~\ref{fig:4}(c)]. This new nodal line gives rise to a new vortex core with opposite topological charge (See Ref. \cite{SI} for the details). Consequently, two additional vortices with opposite topological charges emerge in the far field [Fig.~\ref{fig:4}(d)]. 

The above mechanism can be straightforwardly extended to three-dimensional (3D) systems. To demonstrate this, we calculate the iso-amplitude surfaces ($\left| {{p_{\rm{t}}}} \right| = 0.02$) of the scattered fields for three different scatterer geometries, as shown in Figs.~\ref{fig:4}(e)-\ref{fig:4}(g). In the case of Fig.~\ref{fig:4}(e), the scatterer is a thin plate that approximates the 2D system in Fig.~\ref{fig:3}(a). Accordingly, two vortex tubes emerge, forming a pair of singular lines (white dashed lines) parallel to the $y$-axis [Fig.~\ref{fig:4}(e)]. As the plate dimension is reduced in $y$-direction, the two singular lines bend into two arcs [Fig.~\ref{fig:4}(f)]. When the plate becomes a circular disc, the two singular arcs meet and form a closed singular circle, indicating the generation of a STVR [Fig.~\ref{fig:4}(g)]. As an extension of the 2D STVs, this STVR naturally inherits the propagation stability, and its topological charge can also be reversed via temporal sculpting.

\textit{Experimental verification.\textemdash} To verify the theories for the generation of STVs, we perform acoustic experiments on the 2D platform shown in Fig.~\ref{fig:5}(a) (See Ref. \cite{SI} for the details of experimental set up).  A line source is constructed to generate the incident wave packets. For simplicity, the source amplitude is set to be uniform rather than Gaussian-distributed, which can equally realize the phenomenon. The scattered far fields are scanned along a straight line (dashed red line). The generation of 2D STVs, manipulation of the vortex number, and reversal of the topological charge are shown in the first, second, and third columns of Fig.~\ref{fig:5}(b), respectively. We notice an excellent agreement between the simulation and experimental results. As observed in the panels of the first column, two distinct Y-shaped dislocations are generated in the scattered fields through spatial sculpting. When the plate length increases, corresponding to the panels of the second column, two additional dislocations emerge. The introduction of temporal sculpting further generates two oppositely oriented dislocations, signifying the emergence of opposite topological charges, as shown in the panels of the third column. 

Following similar procedures, we also verify the generation of STVRs on the 3D platform shown in Fig.~\ref{fig:5}(c). The plane sources generate the incident wave packets, which are then scattered by a circular thin disk. The far fields are scanned in a square region on the ($x$,$y$) plane (highlighted in orange). We observe a near-zero-amplitude singularity ring in a snapshot of the far field, signifying the formation of a vortex ring [Fig.~\ref{fig:5}(d)]. Along the $x$- and $y$-directions in the snapshot (dashed white lines in Fig.~\ref{fig:5}(d)), Y-shaped dislocations in the space-time domain are also observed [Fig.~\ref{fig:5}(e)]. The spiral phase profiles near the dislocations (enclosed by dashed squares) further demonstrate the formation of a STVR [Fig.~\ref{fig:5}(f)]. Overall, excellent agreement is observed between the experimental and simulation results, providing solid evidence for the proposed framework of STV and STVR generation.

\textit{Conclusion.\textemdash}In conclusion, we have theoretically and experimentally demonstrated that spatiotemporal coupling provides a robust mechanism for generating stable STVs and STVRs. These topological wave fields exhibit remarkable immunity to the unbalanced diffraction and dispersion, preserving their topological structures over long-distance propagation. Because spatiotemporal coupling is a wave-generic phenomenon, the proposed mechanism is universal and can be generalized to other types of classical waves that may host topological structures, such as electromagnetic waves and surface water waves \cite{peng2022topological,wang2025topological,xiao2026water}. These results provide fundamental insights into the dynamics of spatiotemporal singularities and may find broad applications in ultrafast optics \cite{MA_2022NPho,Abouraddy_2022AOP}, robust OAM transport \cite{Otani_2026CP,ChenZG_2025_NPho}, and vortex ring dynamics within and beyond wave systems \cite{CaiY_2025SA,PhilipH_2023JFM}.

\textit{Acknowledgments.\textemdash}The work described in this paper was supported by grants from National Natural Science Foundation of China (No. 12322416) and the Research Grants Council of the Hong Kong Special Administrative Region, China (Projects No. AoE/P-502/20 and No. JRFS2526-1S11).

\textit{Data availability}---The data that support the findings of this study are available from the corresponding authors upon reasonable request.

\bibliography{reference}




\end{document}